\documentclass[fleqn,usenatbib]{mnras}

\usepackage{mathptmx}

\usepackage[T1]{fontenc}

\DeclareRobustCommand{\VAN}[3]{#2}
\let\VANthebibliography\thebibliography
\def\thebibliography{\DeclareRobustCommand{\VAN}[3]{##3}\VANthebibliography}

\usepackage{graphicx}	
\usepackage{amsmath}	
\usepackage{amssymb}	
\usepackage{subfigure}	
\usepackage[normalem]{ulem}

\title[Galactic normal pulsar population]{Understanding the Galactic population of normal pulsars: A leap forward}

\author[Chakraborty and Bagchi]{ 
Anirban Chakraborty$^{1}$\thanks{E-mail: anirban.chakraborty096@gmail.com} and  Manjari Bagchi$^{2, 3}$\thanks{E-mail: manjari@imsc.res.in} \\
$^{1}$Department of Physics, Indian Institute of Technology Madras, Chennai 600036, India\\
$^{2}$The Institute of Mathematical Sciences, 4th Cross Road, CIT Campus, Taramani, Chennai 600113, India.\\
$^{3}$Homi Bhabha National Institute, Training School Complex, Anushakti Nagar, Mumbai 400094, India
}

\date{Accepted XXX. Received YYY; in original form ZZZ}

\pubyear{2020}

\begin{document}
\label{firstpage}
\pagerange{\pageref{firstpage}--\pageref{lastpage}}
\maketitle

\begin{abstract}

We revisit the population of normal pulsars in the Galactic field in an `evolutionary' approach. By comparing the distributions of various parameters of synthetic pulsars detectable by the Parkes Multibeam Pulsar Survey, the Pulsar Arecibo L-band Feed Array Survey, and two Swinburne Multibeam surveys with those of the real pulsars detected by the same surveys, we find that a good and physically realistic model can be obtained by using a uniform distribution of the braking index in the range of 2.5 to 3.0, a uniform distribution of the cosine of the angle between the spin and the magnetic axis in the range of 0 to 1, a log-normal birth distribution of the surface dipolar magnetic field  with the mean and the standard deviation as 12.85 and 0.55 respectively while keeping the distributions of other parameters unchanged from the ones most commonly used in the literature. We have also replaced the universal `death-line' by a `death-condition' specific to each individual pulsar. We find that our model is better than the most popular model. With our improved model, we predict that an all-sky pulsar survey with phase-I SKA-MID will detect about nine thousand normal pulsars in the Galactic field. Among these pulsars, a considerable number will produce continuous gravitational waves in the operational range of the future ground-based gravitational waves detectors like LIGO A$+$, and in certain cases, the spin-down limit of the gravitational wave strains will be well below the detection sensitivity limit. We also provide a fit for the present-day distributions of the spin periods and 1400 MHz luminosities of the whole normal pulsar population in the Galactic field (which are potentially observable) and those can be used in future population studies under the snapshot approach.

\end{abstract}


\section{Introduction}

Ever since the first discovery in 1967 \citep{hewish_bell}, more than 2500 pulsars have been discovered with a variety of observational characteristics and intrinsic properties. Understanding the birth, evolution, life, and death of pulsars has been of great interest consequently. However, the greatest challenge towards such studies has been the issue of selection biases due to limitations of telescopes and searching algorithms, higher selectivity towards brighter nearby objects, dispersion and scattering of the radio waves in the interstellar medium, etc., in the case of the properties of the observed pulsar samples. Thus, what we observe today is believed to represent only a minuscule portion of a much larger underlying population of pulsars. This strongly motivates the need of statistical studies of pulsar populations. There are two approaches for pulsar population studies. The first is the `snapshot' approach 
where people perform statistical studies on the observed parameters of the known pulsars and use that knowledge to synthesize an entire pulsar population, at that current epoch,  which is then further optimized to match to the observed sample as much as possible. The second one is the `evolutionary' approach where one simulates the distribution of parameters of the newly born pulsars and then model their spin and kinematic evolution, as well as the selection effects, and compare the final distributions of parameters of these synthetic pulsars to the observed sample.

Pulsar population study with the evolutionary approach started almost three decades ago \citep[and references therein]{bwhv82} and was continued by many researchers focusing on various sub-populations of pulsars, e.g., radio and $\gamma$-ray emitting millisecond pulsars \citep{ghf18}, millisecond X-ray pulsars \citep{zlw15}, rotation-powered radio pulsars \citep{fk06}, etc. We set our focus on rotation powered isolated non-recycled radio pulsars in the Galactic field. We use the definition of non-recycled normal pulsars as those having the value of the spin period larger than 0.03 seconds. The last extensive effort to understand various parameters of this population was done in 2006 by \citet{fk06} whose results were later supported by \cite{rl}, \citet{bates}, and many others.

This paper is organised as follows, Section \ref{sec:pop} describes our methodology (including a description of the surveys we use and the comparison strategies adopted) and our analysis results, leading to an improved set of population parameters. We also present the underlying distribution functions for various parameters of the potentially observable pulsar population at the present epoch. We outline results for the normal pulsar population observable by future pulsar surveys like the Square Kilometer Array and discuss the prospects of detection of continuous gravitational waves in various subsections of Section \ref{sec:SKA}. Our main conclusions are summarized in Section \ref{sec:conclusion}.  

\section{Population Studies}
\label{sec:pop}

As we employ population studies within the `evolutionary approach', we first need to simulate a synthetic set of pulsars with chosen birth distributions of their various parameters, e.g., the spin period, the surface magnetic field, the birth locations, the birth velocity, etc, as well as the gravitational potential of the Galaxy. Then, we need to model the motion of the pulsars in the Galaxy to get their final locations, their spin evolution to obtain the present-day spin parameters, as well as model their luminosities. Afterward, to test the validity of these models, we need to compare the properties of the simulated pulsars to those of the observed sample that is affected by observational limitations. Therefore, we further need to pass our synthetic pulsars through the conditions of such limitations as well. For this, we first require to convert pulsar luminosities to flux values at the telescopes using their simulated distances. Then, we need to model the effect of the interstellar medium on the detectability, e.g., scattering and dispersion of the radio waves. For this purpose, we use NE2001 model of the Galactic electron density \citep{ne2001}. Finally, we need to model the telescope sensitivities using the radiometer equation \citep{lk12hpa}.

Detailed step-by-step discussion on this method can be found in \citet{fk06}. We felt the need of redoing the analysis of \citet{fk06} to check whether with a larger sample, the properties of various parameters suggested by \citet{fk06} still remain the same. We also aim to obtain a more physically realistic model. \citet{bates} provided an excellent package `PsrPopPy'\footnote{\url{https://github.com/samb8s/PsrPopPy}.} that can perform evolutionary population synthesis using its module called `evolve'. In addition to the models for various parameters chosen by \citet{fk06}, the `evolve' module has options to use some other models too. In the present work, we use the `evolve' module, but modify it whenever necessary to accommodate our investigations. We also utilise the `dosurvey' module of PsrPopPy to model the detectability of simulated pulsars in various surveys (will be discussed later). For purposes of statistical comparisons in our analysis, we use the {\tt SciPy v1.2.3} package. 

Note that, `PsrPopPy' has another module `populate' that can generate a set of synthetic pulsar population at the present epoch, i.e., it studies the pulsar parameters in a `snapshot' approach. Certain default parameter distributions match with the end results of population studies in the evolutionary approach, as an example, the default luminosity distribution in `populate' is the log-normal distribution obtained by \citet{fk06}. However, in the case of some parameters, there are options to choose distributions different from the default ones. Like `evolve', `populate' can also be used with `dosurvey' to find out how many of the present population of pulsars would be detectable by various surveys and the properties of such detectable population. Due to the ease of use, `populate' interface in PsrPopPy is widely used to understand the performance of various surveys \citep{keane2015,gbncc5,ghrss,HTRU}. We plan to work on the `populate' module and improve it if needed in a future study.

We take the parameters of the real pulsars from the {\tt ATNF Pulsar Catalogue v1.63}\footnote{\url{https://www.atnf.csiro.au/research/pulsar/psrcat/expert.html}} \citep{atnf}. As already mentioned, we restrict ourselves to the study of normal (spin period larger than 0.03 seconds) isolated pulsars as we do not aim to model binary evolution in the present work. We also exclude pulsars in the globular clusters and Magellanic clouds from our study as they are clearly different populations than the ones in the Galactic field.

\subsection{Pulsar surveys included}
\label{sec:surveys}

Before comparing the final parameter distributions of the simulated pulsars with the observed ones, we need to select the pulsars that would be detectable through the surveys that detected the real sample. There are many completed pulsar surveys, and it is not easy to model all those. We select four surveys with the largest pulsar yields at 1.4 GHz, namely the Parkes Multibeam Pulsar Survey (PMPS), the Pulsar Arecibo L-band Feed Array (PALFA) Survey, and two Swinburne Multibeam surveys \citep{pmps_org,palfa_org,swinmb1,swinmb2}. The parameters of these surveys needed to model the detectability of a pulsar with a known value of its flux is given in Table~\ref{tb:surveyList}.

Note that, on the sample of normal real pulsars in the Galactic field discovered by these surveys, we strictly impose the survey sky coverage limits described in Table~\ref{tb:surveyList} to obtain a refined sample of real pulsars and then, exclude the pulsars with missing or anomalous (negative) period derivative values. Ultimately, we are left with a total of 1269 unique real pulsars detected by any one or more of these surveys. This is a larger sample in comparison to \citet{fk06} who used a sample of total 1065 real pulsars. 

There are many other ongoing surveys at lower frequencies like GBNCC, LOTAAS, and AODrift, but we do not use those surveys in the present work, mainly because these surveys are still ongoing. After these surveys are complete, a similar analysis like the present work can be done and will be valuable to understand the population of the non-recycled pulsars better.

\begin{table*}
\begin{center}
  \caption{List of observational survey parameters used for simulating the Pulsar Arecibo L-band Feed Array (PALFA) Survey, Parkes Multibeam Survey (PMPS), and the Swinburne Multibeam pulsar surveys (SWINMB).}
\label{tb:surveyList}
  \begin{tabular}{l c c c}
  \hline
    Parameters & \textbf{PALFA} & \textbf{PMPS} & \textbf{SWINMB} \\ 
    \hline
   	Degradation factor $\beta$ & 1.16 & 1.5 & 1.5\\
	Gain, G (K Jy$^{-1}$ ) &8.2 , 10.4 &0.64 &0.64 \\
	Integration time, t$_{obs}$  (s) & 268 & 2100 & 265\\
	Sampling interval, t$_{samp}$  (ms) & 0.065 & 0.250 & 0.125 \\
	Receiver Temperature, T$_{rec}$ (K) & 24 & 25 & 25 \\
	Centre frequency, f (MHz)& 1375 & 1374 & 1374  \\
	Bandwidth, BW (MHz)& 322 & 288 & 288 \\
	Channel width, ${\Delta}$f (MHz) & 0.34 & 3 &3\\
	Number of polarizations, n$_{p}$ & 2 & 2 & 2 \\
	Beam FWHM (arcmin) & 3.35 & 14.0 & 14.0 \\
	Min. RA ($^{\circ}$)& 0 & 0 & 0  \\
	Max. RA ($^{\circ}$)& 360 & 360 & 360 \\
	Min. Dec ($^{\circ}$)& 0&-90 &-90  \\
	Max. Dec ($^{\circ}$)& +38& +90 & +90 \\
	Galactic longitude coverage & +32$^{\circ}$ $\leq$ $l$ $\leq$ +77$^{\circ}$  & -100$^{\circ}$ $\leq$ $l$ $\leq$ +50$^{\circ}$ & -100$^{\circ}$ $\leq$ $l$ $\leq$ +50$^{\circ}$\\
	Galactic latitude coverage & $|b|$ $\leq$ 5$^{\circ}$  & $|b|$ $\leq$ 5$^{\circ}$ & 5$^{\circ}$ $\leq$ $|b|$ $\leq$ 30$^{\circ}$\\
	Detection S/N & 11.3& 9.0 &9.0\\
	Gain Pattern &Airy&Gaussian & Gaussian\\
 	\hline 
  \end{tabular}
\end{center}
\end{table*}

\subsection{Simulation and Model Comparison Strategy}
\label{sec:sim_procedure}

In order to compare the properties of synthetic pulsars to those of the real ones, we make use of the two-sample Kolmogorov Smirnoff test. This nonparametric test compares the cumulative distribution function of two data sample sets and returns a D-statistic, which can then be used to infer the probability $f$ that the two data samples are drawn from the same underlying distributions. For each simulated and `detected' set of pulsars, the distributions of the Galactic longitudes ($l$), Galactic latitudes ($b$), dispersion measures (DM), flux densities at 1400 MHz ($S_{1400}$), spin periods ($P$), and spin period derivatives ($\dot{P}$) are independently compared with the respective distributions for the real sample, and following \citet{bwhv82} we assign an overall goodness of fit metric $F$ for the simulation model, as follows:

\begin{equation}
\label{eq:FOM}
F=\log_{10} \bigg[ f_{l} \times f_{b} \times f_{\rm DM} \times f_{S_{1400}} \times f_{P} \times f_{\dot{P}}  \bigg] ~.
\end{equation}

Note that $F$ is a negative quantity as the probabilities ($f$) for various parameters are always less than one. The model that produces the minimum absolute value of $F$ can be considered as the best one. For computing the KS test probabilities of a particular simulation model, we use all the simulated pulsars generated, i.e., we use $ \sim 1269 \times$ $N_{\rm trials}$ synthetic pulsars, where $N_{\rm trials}$ is the number of realisations ($= 50$) carried out for that particular model. However, in the sample of the real pulsars, there are instances where the values of $S_{1400}$ are not available in the catalogue. In that case, we follow the same strategy as was adopted by \citet{fk06}, i.e., only while comparing the distribution of $S_{1400}$ between the simulated and real samples, we use a reduced sub-sample of real pulsars excluding the ones with missing $S_{1400}$ values.

\subsection{Results from FK06 optimal model}
\label{sec:fk06_default}

\begin{figure*}
\begin{center}
\subfigure[Model A (see Table \ref{tb:FK06Results})]{\label{subfig:ModelA}\includegraphics[width=\textwidth,height=0.45\textheight]{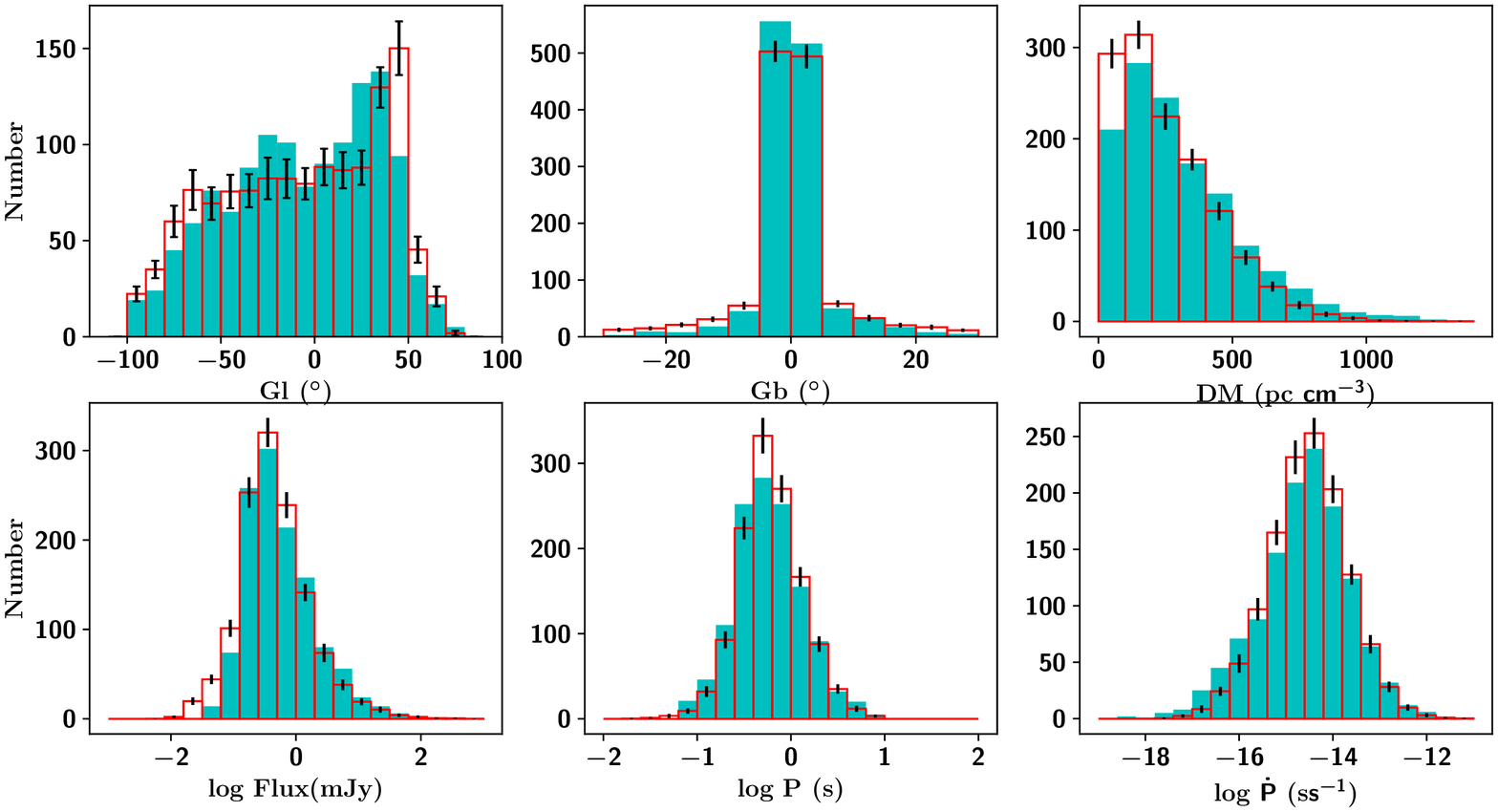}}
\subfigure[Model B (see Table \ref{tb:FK06Results})]{\label{subfig:ModelB}\includegraphics[width=\textwidth,height=0.45\textheight]{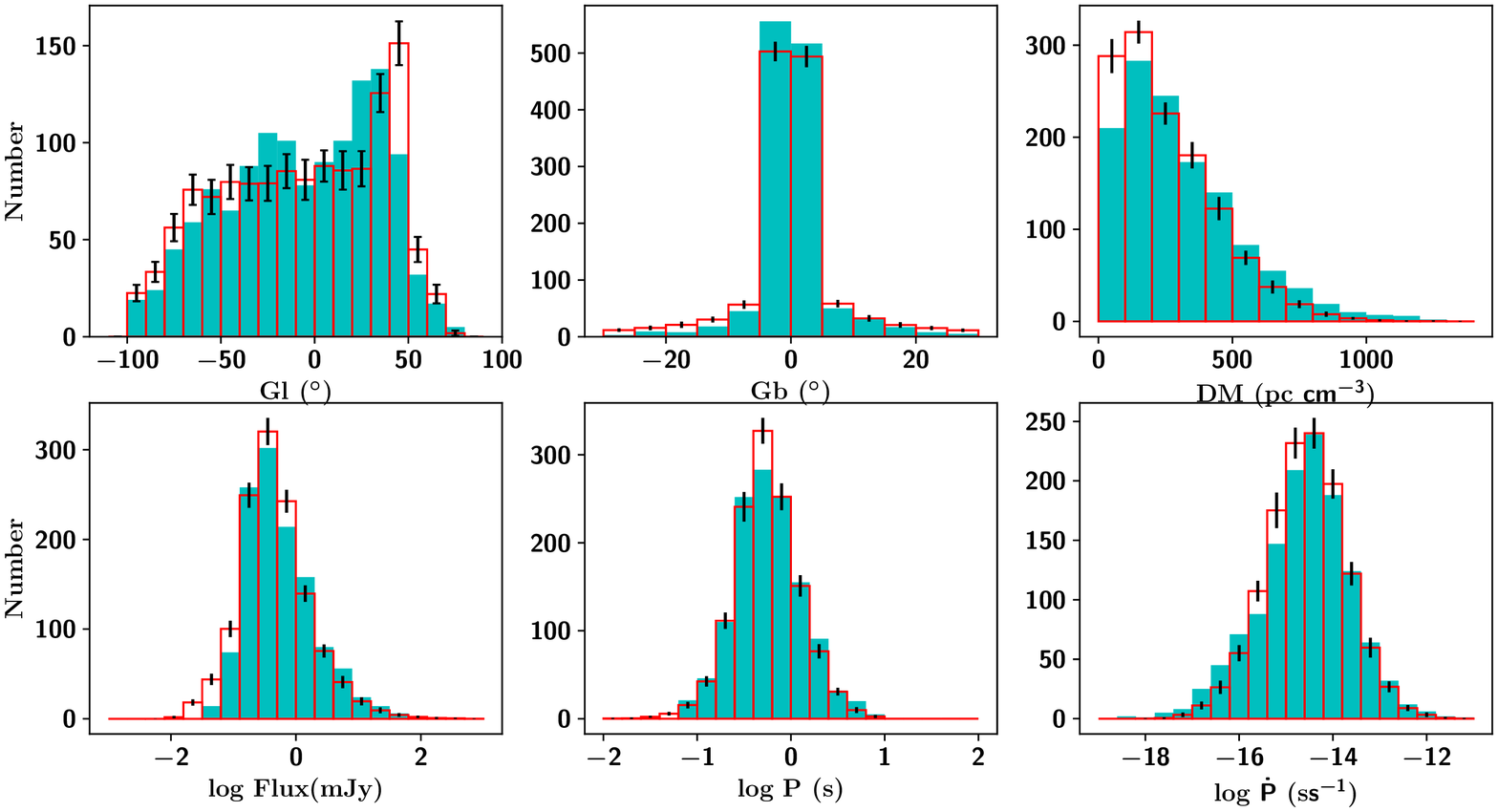}}
 \end{center}
\caption{Distributions of Galactic longitudes, Galactic latitudes, dispersion measures, flux densities at 1.4 GHz, spin periods, and spin period derivatives for the synthetic pulsars (red bar plots) simulated for Model-A and Model-B as described in the text and those of real observed pulsar (solid turquoise-blue histograms). For the red bar plots (simulations), the number of pulsars in each bin is the average over 50 realizations of the models.}
\label{fig:FK06}
\end{figure*}

We start by constructing a pulsar population model (Model-A) based on the distributions and corresponding best-fit parameters found by \citet{fk06}, which has been outlined in Table \ref{tb:FK06Model}. Figure \ref{subfig:ModelA} shows the distributions of various parameters of the synthetic and real pulsars and the first column of Table \ref{tb:FK06Results} reports the KS test results on various parameters as well as the goodness of fit $F=-27.8415$. We see that there is a reasonably good agreement between the properties of the detected synthetic pulsars and the real ones. However, we aim to obtain a model that is better than this Model-A, both physics wise and in mimicking the parameter distributions of the observed sample. Our steps to achieve this goal are described in the subsequent sub-sections.

\begin{table}
\caption{Parameters for the best model of \citet{fk06}, which we call as Model-A.}
\vskip -0.1cm
\centering
\label{tb:fk06model}
\begin{tabular}{l|l} \hline \hline
   Model Parameters &  Values\\
\hline
   Pulsar Age Distribution &  Uniform      \\
    $t_{max}$ &  1 Gyr      \\
   \hline
   Radial Distribution &  Yusifov $\&$ K{\"u}{\c{c}}{\"u}k      \\
  $R_{1}$ &  $0.55$ kpc      \\
  a &  $1.64$      \\
   b &  $4.01$      \\
  \hline
  Birth Height Distribution & Exponential \\
   $\left\langle z_{0} \right\rangle$ &  $50 $ pc \\ 
   \hline
  Birth Velocity Distribution & Exponential \\
  $\left\langle v_{3D} \right\rangle$ &  $380$ km $s^{-1}$ \\ 
   $\left\langle v_{1D} \right\rangle$ &  $180$ km $s^{-1}$ \\ 
   \hline  
   Birth Spin Period Distribution & Normal \\
  $\left\langle P_0 \right\rangle$ &  $300$ ms  \\
   $ \sigma_{P_0} $ &  $150$ ms  \\
  \hline 
   Magnetic Field Distribution & Log-normal (in the logarithm to the base 10)\\
  $\left\langle log (B/G) \right\rangle$ &  12.65  \\
   $ \sigma_{log B} $ &  0.55 \\ 
  \hline
   Rotational Alignment &  \\
    $\alpha $ &  $90\degr$  \\
   \hline 
   Luminosity Model & $P$-$\dot{P}$ power law \\
   $L_{0} $ &  $0.18$ mJy $kpc^{2}$  \\
   $ \epsilon_{P}$  &  $-1.5$  \\
   $ \epsilon_{\dot{P}} $ &  $0.5$  \\ 
   $ \sigma_{L_{corr}} $ &  $0.8$ \\   
   \hline \hline
\label{tb:FK06Model}
\end{tabular}
\end{table}

\begin{table*}
\caption{KS test probabilities obtained on comparing the distributions of the parameters of the real pulsars with those of the
simulated ones for models A, B, and C as described in the text. The results are computed using 50 realisations of the particular model. Here `$\alpha$' refers to the angle between the spin and the magnetic axis and `$n$' refers to the braking index. Model-A is the model chosen by \citet{fk06}. }
\vskip -0.1cm
\centering
\begin{tabular}{l c c c c} \hline \hline
   Parameters & Model A  &  Model B  & Model C & \\
   \hline
   & $n$ = 3  &  2.5 $<$ $n$ $<$ 3 & 2.5 $<$ $n$ $<$ 3& \\
    &  $\alpha$ = 90\degr & $\alpha$ = 90\degr & uniform distribution of $\cos \alpha$ & \\
\hline
  Galactic longitude $l$ (deg) & $0.00027$ &$0.00026$   &$0.00104$  &  \\
  Galactic latitude  $b$ (deg) & $0.01966$ &$0.00682$   &$5.148\times 10 ^{-5}$ &  \\
  Dispersion Measure $DM$ (pc/cc)& $5.966 \times 10 ^{-11}$ &$1.057 \times 10 ^{-9}$ & $2.512 \times 10 ^{-14}$ & \\ 
Flux Density $S_{1400}$  & $3.566 \times 10 ^{-6} $ & $1.198 \times 10 ^{-5} $ & $9.736 \times 10 ^{-6}$   \\ 
   Period $P$ (s) & $ 0.00030$ & $0.02202$ & $ 9.161 \times 10 ^{-9} $&  \\
  Period Derivative $\dot{P}$ (s s$^{-1}$)& $ 0.00041 $ & $0.00510$ & $2.558 \times 10 ^{-19}$  \\
  \hline    Overall Goodness of Fit Metric $ F $& -27.8415 & -23.5934 & -52.5110   \\
   \hline \hline
\label{tb:FK06Results}
\end{tabular}
\end{table*}

\subsection{Steps to a more realistic model}

\citet{rl} proposed that for a small deviation of the dipole nature of the magnetic field, the spin-down equation can be written (in CGS system of units) as: 
\begin{equation}
P~^{n-2}\dot{P} = a \, k \, B^2 sin^2 \alpha ~,
\label{eq:Psrpoppy_SpinDown}
\end{equation}
where $a$ is a constant in the unit of $s^{n-3}$, $\alpha$ is the angle between the spin and the magnetic axis, $n$ is the pulsar braking index, $B$ is the magnetic field at the surface of the neutron star. The constant $k$ is equal to $(8/3) \, \pi^2 R^{6} I^{-1} c^{-3}$ where $I$ is the moment of inertia and $R$ is the radius of the neutron star. For the canonical values for neutron stars, i.e., $R=10^6~{\rm cm}$, $I= 10^{45} ~{\rm gm~cm^2}$, $k$ becomes 9.76 $\times 10^{-40} ~{\rm cm ~gm^{-1} ~ s^{-1}}$. For $n=3$, the constant $a$ takes the value of 1, and equation (\ref{eq:Psrpoppy_SpinDown}) becomes the standard spin-down equation for the dipolar magnetic field.

It has been proposed \citep{bh91} that radio pulsars cannot emit if $B \le 0.17 \times 10^{12}  P^2  $ where $B$ is in Gauss and $P$ is in seconds, and the condition can be called the `death-condition'. 

Using Equation (\ref{eq:Psrpoppy_SpinDown}), we find that the `death-condition' for the pulsars becomes
\begin{equation}
\frac{B}{P^2}=\frac{\sqrt{P^{n-2} \dot{P}}}{\sqrt{a \, k} \sin \alpha \, P^2} \leq 0.17 \times 10^{12} ~{\rm G~ s^{-2}} ~. 
\label{eq:My P-Pdot deathline}
\end{equation}
For $n=3$ and $\alpha = 90 \degr$, the above equation becomes 
\begin{equation}
\frac{B}{P^2}= \frac{\sqrt{P~\dot{P}}}{\sqrt{k} \, P^2} \leq 0.17 \times 10^{12} ~{\rm G~ s^{-2}} ~. 
\label{eq:BPdeathline}
\end{equation} where the equality represents the standard `death-line' equation \citep{bwhv82}. For a phenomenological model, where the magnetic field deviates slightly from the dipolar nature, the value of the constant $a$ can be taken as 1 as done by \citet{rl}. However, we need to remember the existence of this constant, as otherwise the dimensionalities of the two sides of equation. (\ref{eq:Psrpoppy_SpinDown}) would be different when $n \neq 3$.

From Equation (\ref{eq:My P-Pdot deathline}), it is clear that if we use a distribution of $n$ and $\alpha$, then each pulsar has its own characteristic `death-condition', not a universal `death-line' for all pulsars, as was the case for \citet{bwhv82} or \citet{fk06} who worked with a particular value of the braking index ($n=3$) and the inclination angle ($\alpha = 90 \degr$) for all synthetic pulsars.

\subsubsection{A model for the braking index distribution}
\label{sec:improve_ndist}

\citet{fk06} in their analysis had assumed a braking index of $n$ = 3 for all synthetic pulsars. However, for a good number of real pulsars, the measured values of $n$ differ from 3. Moreover, \citet{rl} showed that for their optimal model, a uniform distribution of $n$ in the range of 2.5 $-$ 3.0 provided a better agreement (see Table 3 in their paper). We test this hypothesis by adopting such a distribution for $n$ while keeping all other model parameters the same as those in Model-A. We call this new model as Model-B.

Figure \ref{subfig:ModelB} shows the distributions of various parameters of the synthetic and real pulsars and the second column of Table \ref{tb:FK06Results} reports the KS test results on various parameters as well as the goodness of fit $F=-23.5934$ for Model-B. We see better agreement between the simulated and observed distributions, particularly in the $P$ and $\dot{P}$ distributions for Model-B in comparison to Model-A. Therefore, we decide to choose a uniform distribution of braking index in the range of 2.5 $-$ 3.0 and look for improvements in the choice of other parameters.

\subsubsection{A distribution for the angle between the spin and the magnetic axis}
\label{sec:improve_alphadist}

\begin{figure*}
\begin{center}
\includegraphics[width=\textwidth,height=0.45\textheight]{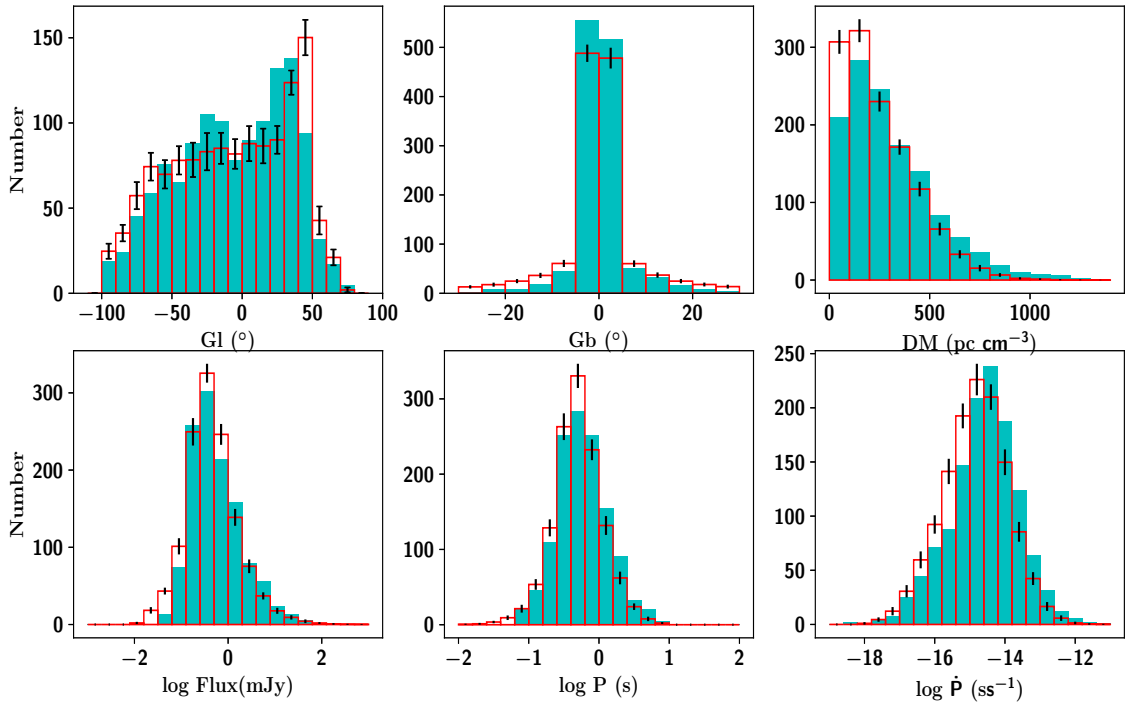}
\end{center}
\caption{Distributions of Galactic longitudes, Galactic latitudes, dispersion measures, flux densities at 1.4 GHz, spin periods, and spin period derivatives for the synthetic pulsars (red bar plots) for model-C as described in the text and those of real observed pulsar (solid turquoise-blue histograms). For the red bar plots (simulations), the number of pulsars in each bin is the average over 50 realizations of the model.}
\label{fig:ModelC}
\end{figure*}

In order to obtain a more realistic picture of the pulsar population, we deviate from \citet{fk06} assumption of the orthogonal rotator ($\alpha$ = 90$^{\circ}$). We select a uniform distribution of $x = \cos \alpha$ in the range of 0 to 1, keeping all other parameters the same as Model-B. We call this new model Model-C. The results of this model is given in the third column of Table \ref{tb:FK06Results}. However, instead of better (smaller absolute value) $F$, we get a much worse result in comparison to models A and B (both are for orthogonal rotator). This fact is also supported by Figure \ref{fig:ModelC} where  distributions of various parameters of the synthetic and real pulsars are shown. It is clear that the highest discrepancies are in the distributions of $P$ and $\dot{P}$. This result inspired us to check whether the birth distributions of the surface magnetic field (which is used in calculating the spin period and period derivative) need to be improved.

\subsubsection{Improved model for the birth distribution of the surface magnetic field}
\label{sec:improve_bdist}

The magnetic field $B$ at the surface of the neutron star is expressed as a function of $P$, $\dot{P}$ and $\alpha$ as seen in equation (\ref{eq:Psrpoppy_SpinDown}) and the assumption of the orthogonal rotator ($\alpha = 90^{\circ}$) gives the lower limit of $B$. So, it is reasonable to expect that the deviation from the orthogonal rotator model should be accompanied by a change in the birth distribution of $B$. Keeping the distribution type log-normal (in the logarithm to the base 10), we vary the values of $ \mu_{log B} $ and $ \sigma_{log B} $ over a wide range to see whether we can obtain a better agreement between the parameters of the simulated and the real pulsars. We cover the parameter space spanned by 12.35 $\leq$ $\mu_{log B}$ $\leq$ 13.15 (in steps of 0.1) and 0.40 $\leq$ $\sigma_{log B}$ $\leq$ 0.70 (in steps of 0.05). For each grid point ($\mu_{log B}$,$\sigma_{log B}$), we performed 50 realisations of the particular simulation model and determined its goodness of fit metric using equation (\ref{eq:FOM}) as before. 

\begin{figure*}
\begin{center}
\includegraphics[width=\textwidth,height=0.5\textheight]{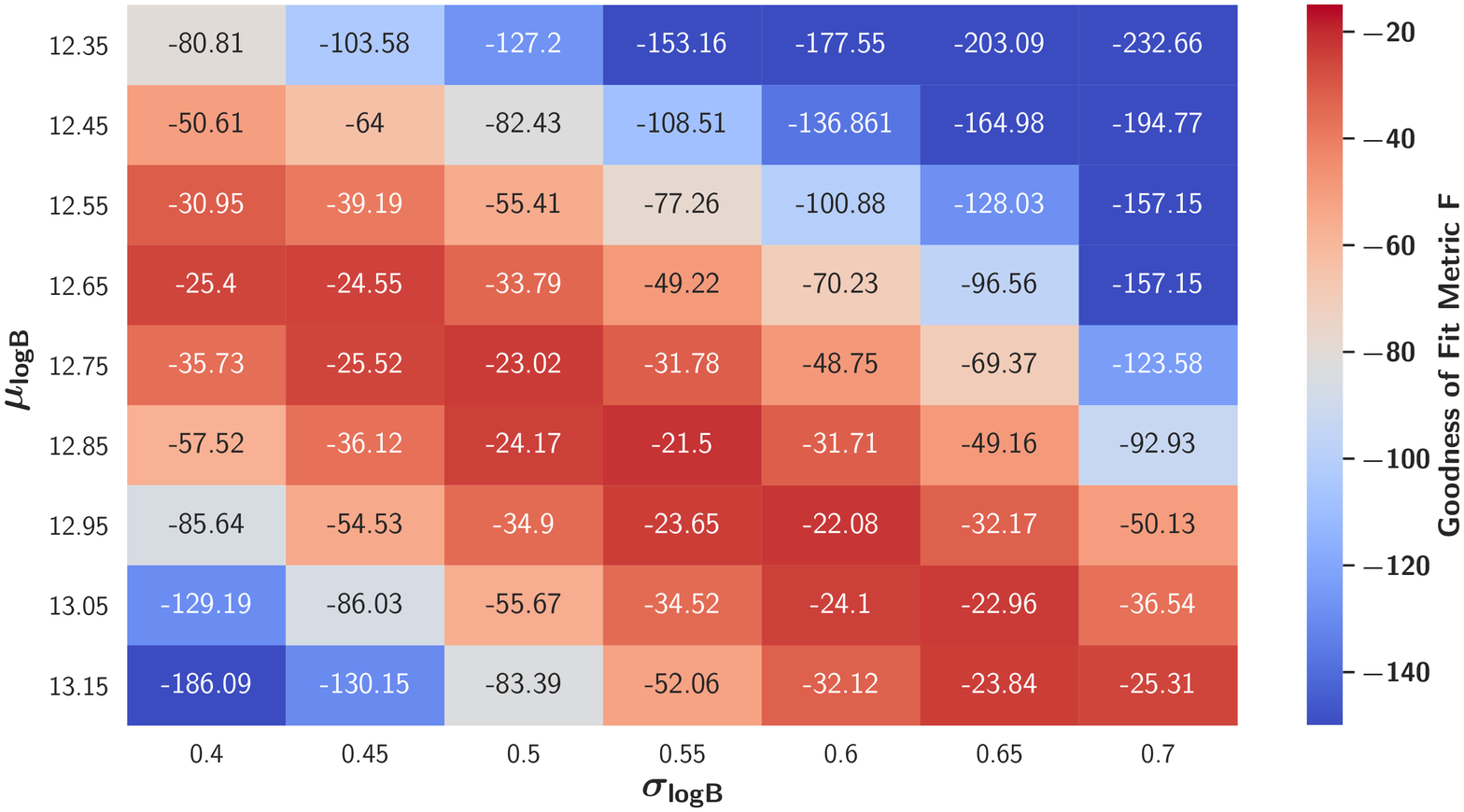}
\end{center}
\caption{The variation of the computed goodness of fit value F (see Equation \ref{eq:FOM})
over the $\mu_{logB} - \sigma_{logB}$  parameter space. For each grid point, 50 realisations of the particular model were carried out and the corresponding goodness of fit value F is mentioned.}
\label{fig:heatmap}
\end{figure*}

A heat map showing the variation of the computed goodness of fit value $F$ over the scanned $\mu_{logB} - \sigma_{logB}$ parameter space is shown in Figure \ref{fig:heatmap}. It is clear that there exists a broad range of values of $\mu_{logB}$ and $\sigma_{logB}$ for which the simulated pulsar parameter distributions agree reasonably well with the observed sample. We identify the `best' combination of parameters to be those that corresponds to the case where we obtained the maximum value of $F=-21.5$, i.e., $\mu_{log B},\sigma_{log B}$ = $12.85~,~0.55$. We call this model as Model-D where other than the magnetic field, everything else is the same as in Model-C.

\begin{figure*}
	\centering
	\includegraphics[width=\textwidth,height=0.45\textheight]{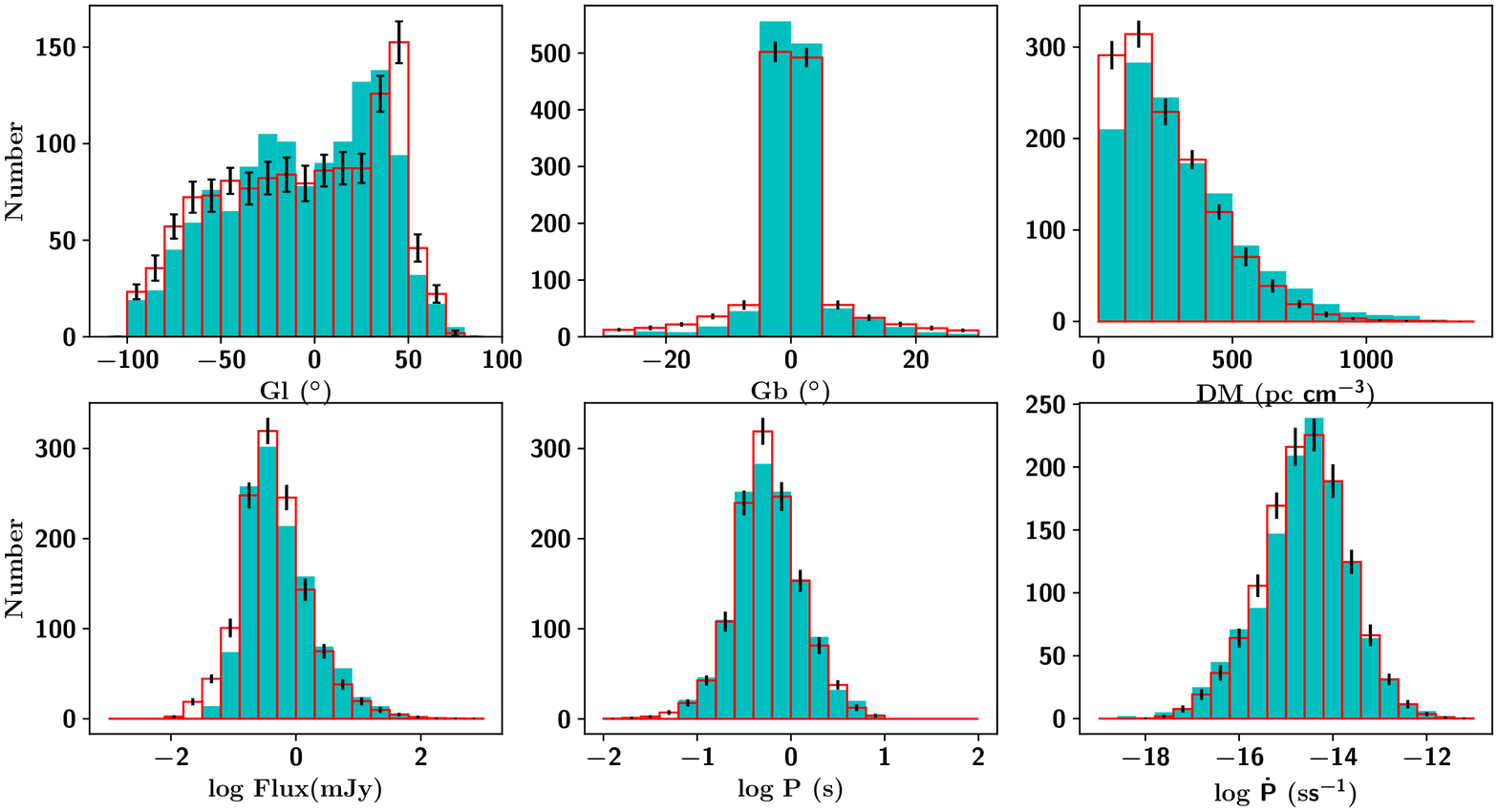}
	\caption{Distributions of Galactic longitudes, Galactic latitudes, dispersion measures, flux densities at 1.4 GHz, spin periods, and spin period derivatives for the synthetic pulsars (red bar plots) simulated for Model-D as described in the text and those of real observed pulsar (solid turquoise-blue histograms). For the red bar plots (simulations), the number of pulsars in each bin is the average over 50 realizations of the model.}
	\label{fig:ModelD}
\end{figure*}

\begin{figure*}
	\centering
	\includegraphics[width=0.75\textwidth,height=0.45\textheight]{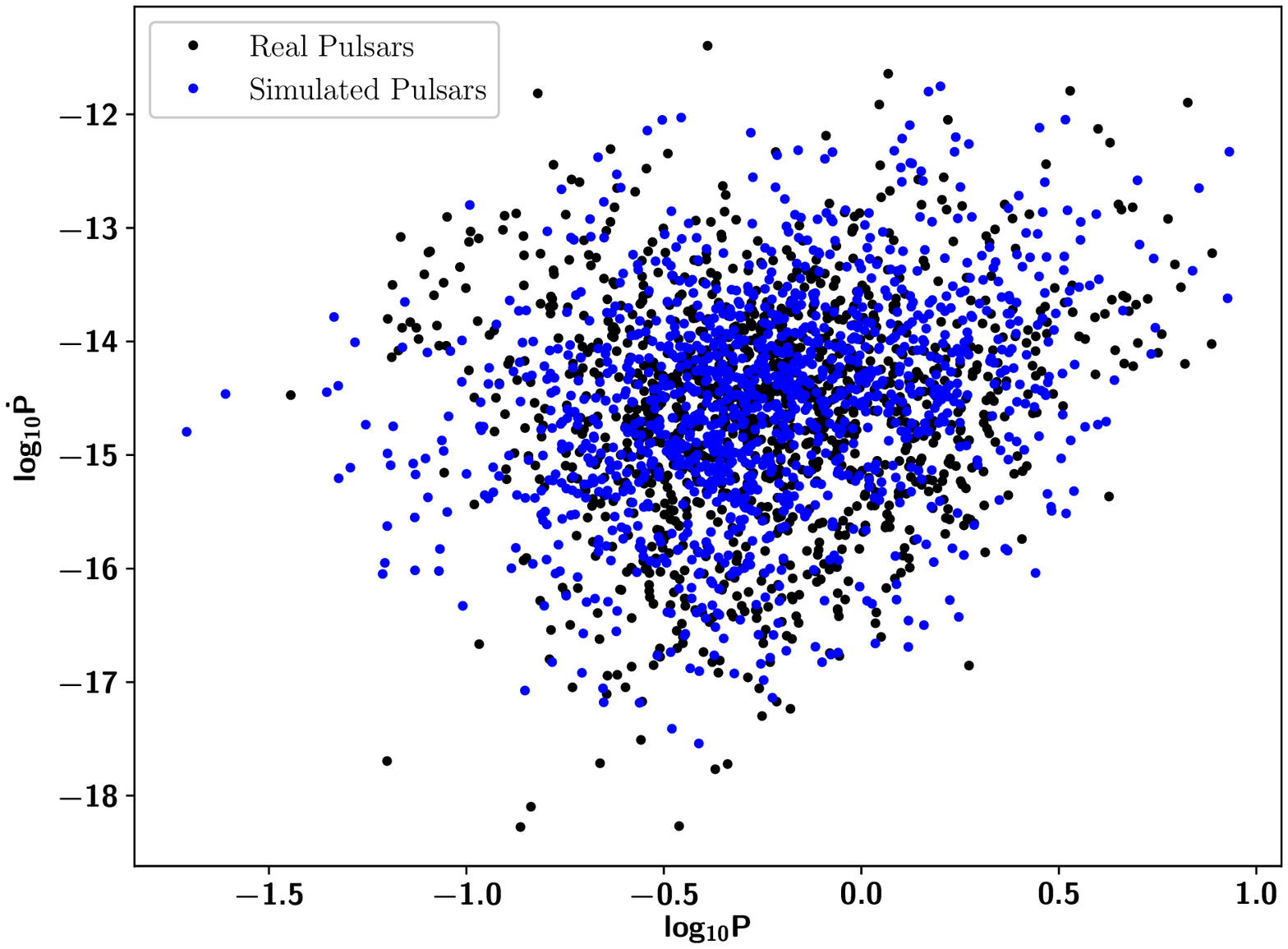}
	\caption{An example $P$ - $\dot{P}$ diagram generated for the detectable pulsars (blue dots) in a single realization of Model-D. For comparison, we also show the real observed pulsars (black dots) on the same diagram}
	\label{fig:ModelD_PPdot}
\end{figure*}

The KS Test probabilities, obtained by comparing the distributions of parameters of the simulated pulsars following Model-D with those of the real pulsars, are reported in Table \ref{tb:modelDCompare}. It is clear that statistically, our simulation model shows good agreement with the real pulsars, even better than Model-A (the model used by \citet{fk06}). The distribution of various pulsar parameters for a population generated according to Model-D is shown in Figure \ref{fig:ModelD}. An example $P$-$\dot{P}$ diagram generated for a single realization of model-D is shown in Figure \ref{fig:ModelD_PPdot} where $P$, $\dot{P}$ values for real pulsars (used in this work) are also shown.

\begin{table}
\caption{KS test probabilities obtained by comparing the parameters of the real pulsars with those of the simulated and detected ones corresponding to Model D. The results are computed using 50 realisations of the model.}
\vskip -0.1cm
\centering
\begin{tabular}{|l|c|} 
 \hline \hline
   Parameters  &    Model D \\ \hline
   \hline
	\hline
   Galactic longitude $l$ (deg) & $0.00056$  \\
   Galactic latitude  $b$ (deg) & $0.00438$ \\
   Dispersion Measure DM (pc/cc)& $4.70 \times 10 ^{-10}$  \\ 
 Flux Density $S_{1400}$  & $6.99 \times 10 ^{-6} $   \\ 
 Period $P$ (s) & $ 0.1021$ \\
  Period Derivative $\dot{P}$ (s s$^{-1}$)& $ 0.3743 $ \\
  \hline  Overall Goodness of Fit Metric $ F $& -21.505  \\
   \hline \hline
\label{tb:modelDCompare}
\end{tabular}
\end{table}

\subsection{Underlying distributions of pulsar population}
\label{sec:underlying_pop}

\begin{figure*}
	\centering
	\begin{center}
\subfigure[1400 MHz Pulsar Luminosity]{\includegraphics[width=0.32\textwidth,height=0.22\textheight]{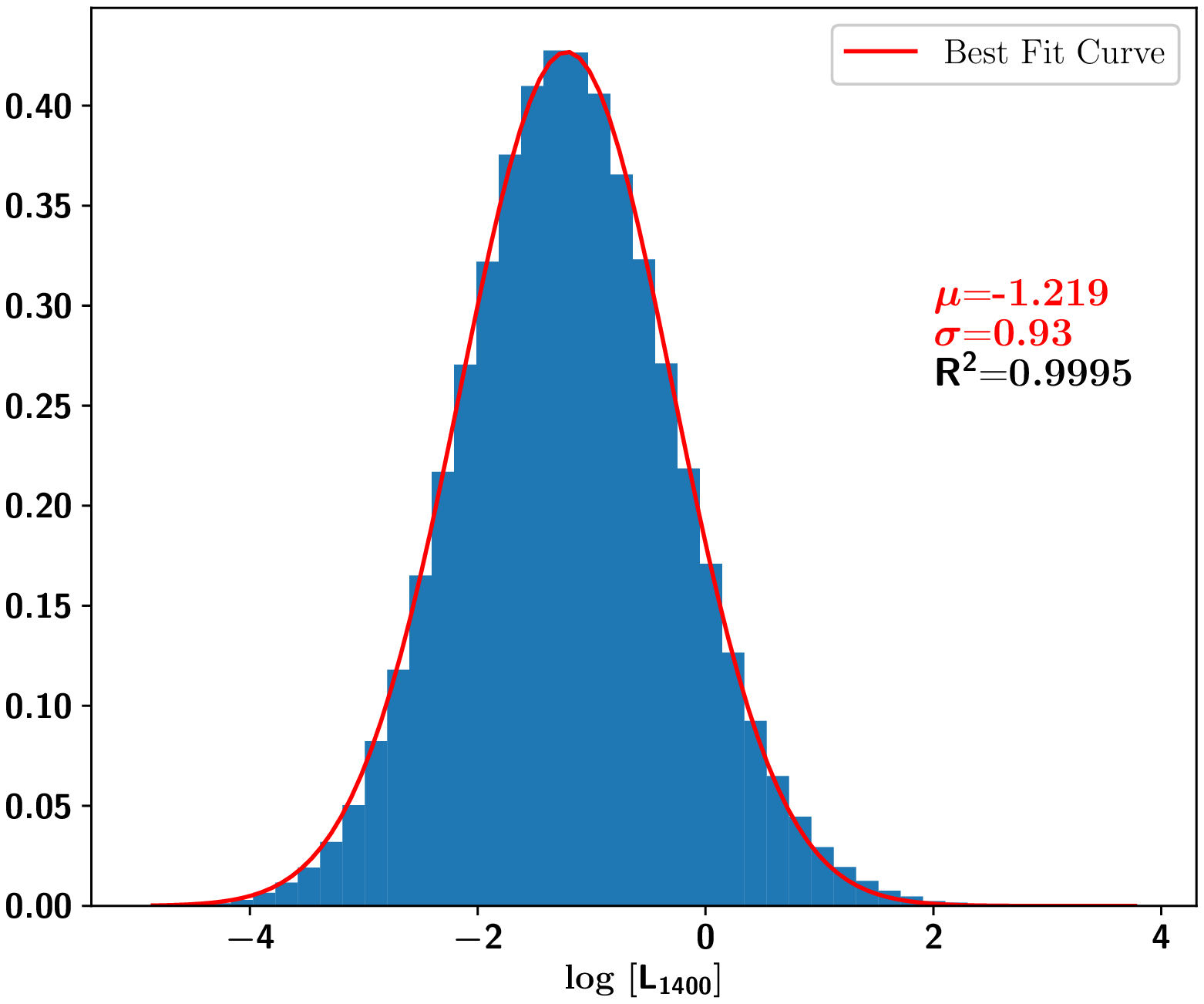}}
\subfigure[Spin Period ]{\includegraphics[width=0.32\textwidth,height=0.22\textheight]{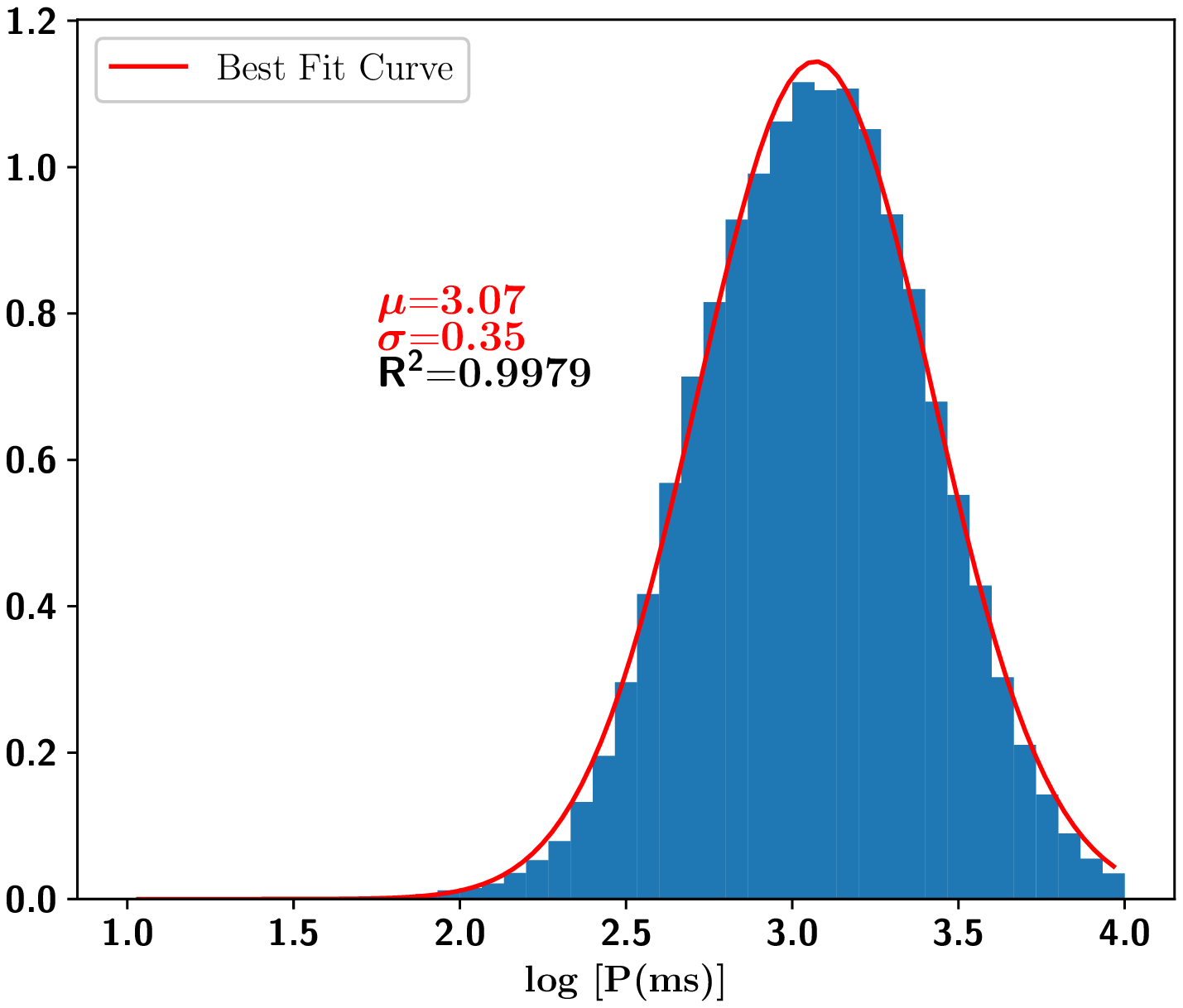}}
\subfigure[Spin Period Derivative]{\includegraphics[width=0.32\textwidth,height=0.22\textheight]{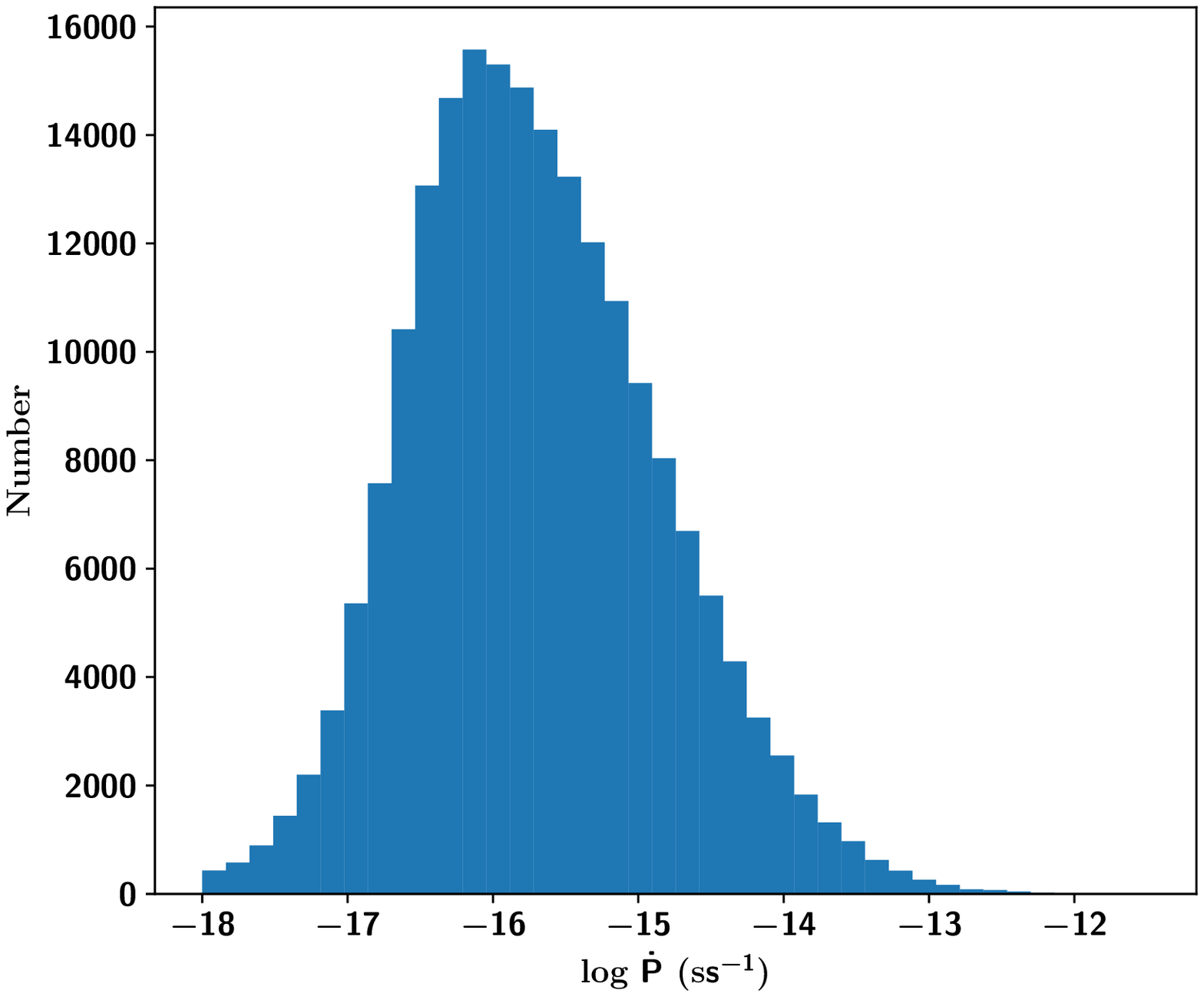}}
\end{center}
\caption{Underlying distributions of pulsar luminosities at 1.4 GHz, spin periods, and spin period derivatives in a single realization of Model D. The solid red line in each case represents the best-fit curve for the corresponding distribution.}
\label{fig:underlying}
\end{figure*}

Using Model-D, we investigate whether the underlying distribution of some important parameters like $P$, $\dot{P}$, and $L_{1400}$ for the whole normal pulsar population in the Galaxy (which are radio-loud and potentially observable) can be fitted with a model that can be used for population studies under the `snapshot' approach. We find that the underlying $L_{1400}$ distribution is well fit by a log-normal function (in the logarithm to the base 10) with mean $\mu_{log L} =-1.219$ ($L$ = $0.06$ mJy kpc$^2$) and standard deviation $\sigma_{log L}$ = $0.93$. Note that \citet{fk06,rl} also found log-normal distribution for $L_{1400}$, but \citet{fk06} had found $\mu_{log L} = -1.1 $ and $\sigma_{log L} = 0.9 $. We also find that a log-normal distribution of $P$ (in units of milliseconds) with $\mu_{logP} = 3.07$ and $\sigma_{logP} = 0.35$ fits the underlying distribution well.

The distributions and their corresponding best-fit curves have been shown in Figure \ref{fig:underlying}.

\section{Pulsar Science in Square Kilometer Array (SKA) Era}
\label{sec:SKA}

The Square Kilometre Array (SKA)\footnote{\url{https://www.skatelescope.org/}}, is expected to become operational in the coming decade. It is believed that SKA will advance pulsar science by leaps and bounds and open innumerable avenues for a wide range of scientific investigations mostly by discovering many new pulsars \citep{keane2015}. 

In this section, we assess the pulsar discovery capabilities with SKA Phase-1 within the framework of our improved Model-D. We focus only on using SKA-MID telescope and simulate a potential all-sky survey with the same, using the default survey template file from {\tt PsrPopPy} and accounting for the resources as documented in the design document\footnote{\url{https://www.skatelescope.org/wp-content/uploads/2012/07/SKA-TEL-SKO-DD-001-1_BaselineDesign1.pdf}.}. The parameters for the all-sky pulsar search survey using SKA-MID is listed out in Table \ref{tb:SKAsurveyList}.

\begin{table}
\begin{center}
  \caption{List of assumed survey parameters used for simulating the hypothetical SKA-MID pulsar survey.}
\label{tb:SKAsurveyList}
  \begin{tabular}{l c}
  \hline
    Parameters & \textbf{SKA-MID}\\ 
    \hline
   	Degradation factor $\beta$ & 1.0\\
	Gain, G (K Jy$^{-1}$ ) &4.5 \\
	Integration time, t$_{obs}$  (s) & 600\\
	Sampling interval, t$_{samp}$  (ms) & 0.05 \\
	Receiver Temperature, T$_{rec}$ (K) & 20 \\
	Centre frequency, f (MHz)& 1400 \\
	Bandwidth, BW (MHz)& 300 \\
	Channel width, ${\Delta}$f (MHz) & 0.020 \\
	Number of polarizations, n$_{p}$ & 2 \\
	Beam FWHM (arcmin) & 4.5 \\
	Min. RA ($^{\circ}$)& 0  \\
	Max. RA ($^{\circ}$)& 360 \\
	Min. Dec ($^{\circ}$)& -90\\
	Max. Dec ($^{\circ}$)& +30 \\
	Galactic longitude coverage & -180$^{\circ}$ $\leq$ $l$ $\leq$ +180$^{\circ}$\\
	Galactic latitude coverage & $|b|$ $\leq$ 90$^{\circ}$ \\
	Detection S/N & 9.0\\
	Gain Pattern &Gaussian\\
 	\hline
  \end{tabular}
\end{center}
\end{table}

\subsection{Survey yields from all-sky SKA-MID pulsar search}
\label{subsec:SKA_yields}

\begin{figure*}
	\centering
	\includegraphics[width=\textwidth]{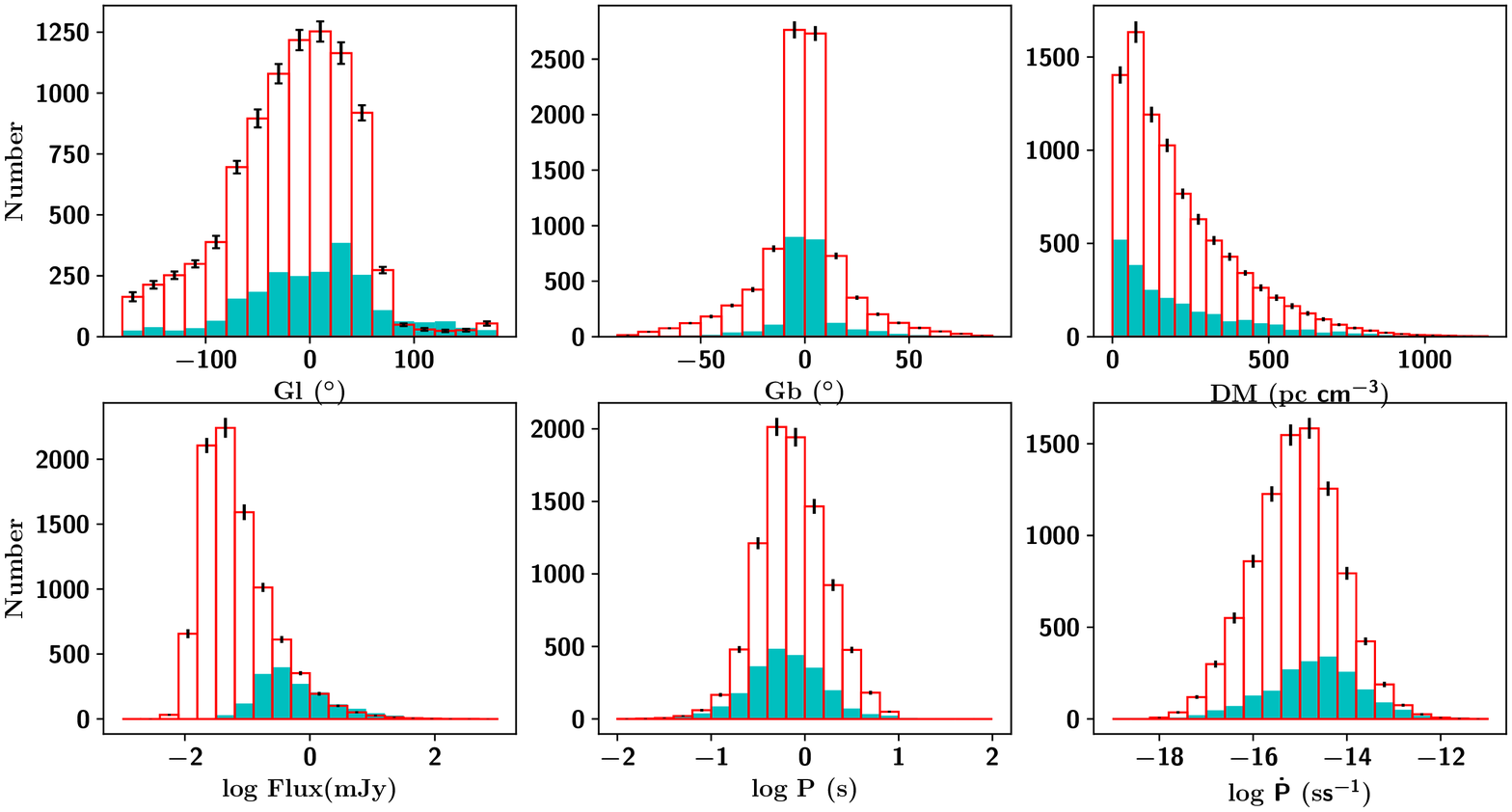}
	\caption{Distributions of Galactic longitudes, Galactic latitudes, dispersion measures, flux densities at 1.4 GHz, spin periods, and spin period derivatives for the synthetic pulsars (red bar plots) detected in the simulated SKA-MID survey described in Table \ref{tb:SKAsurveyList}. For the synthetic pulsar set, the number of pulsars in each bin is the average over 50 realizations of Model-D. The filled turquoise-blue histograms depict the corresponding distributions for the whole sample of normal real pulsars in the Galatic field (around 2300 pulsars) that have been detected so far using any terrestial telescope.}
	\label{fig:SKA_Detections}
\end{figure*}

From 50 realisations of Model-D, we deduce that the hypothetical SKA-MID pulsar survey will be able to detect about 9000 normal pulsars in the Galactic field (including re-detection of already detected pulsars that fall in its visible sky region). 

The distribution of the different pulsar parameters of the detectable population from such a survey is shown in Figure \ref{fig:SKA_Detections}. This figure also shows the distributions of the same parameters of all the known pulsars (about 2300) in the Galactic field. Note that, this sample of real normal pulsars in the Galactic field is different (larger) than the original sample (outcome of only four surveys) that we used to constrain our model, as here our motivation is to see how SKA will enhance the present population. It is clear that SKA-MID will significantly increase the population of long period pulsars ($P$ $\geq$ 1s) and also provide a wholesome census of fainter pulsars, which will be useful in further constraining the luminosity distribution function of normal pulsars.

\begin{figure*}
	\centering
	\includegraphics[width=0.80\textwidth,height=0.40\textheight]{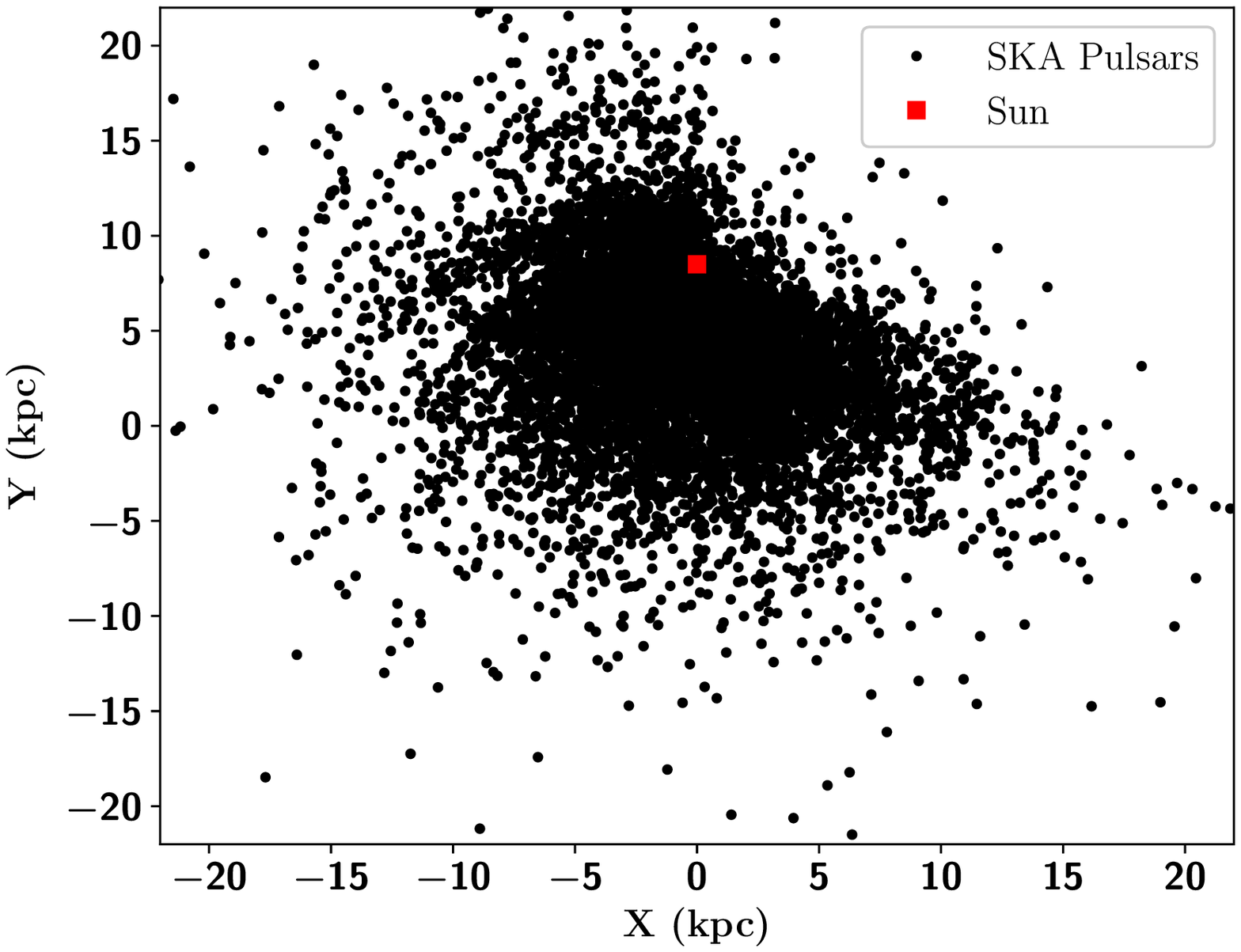}
	\caption{Locations of the simulated normal pulsars that are potentially detectable by the considered hypothetical SKA-MID pulsar survey, projected onto the Galactic plane from a single realisation of Model-D. The Galactic centre is at the origin while the Sun (shown as a red square) is located at (0, 8.5) kpc.}
	\label{fig:SKA_XY}
\end{figure*}

\begin{figure*}
	\centering
	\includegraphics[width=0.80\textwidth,height=0.45\textheight]{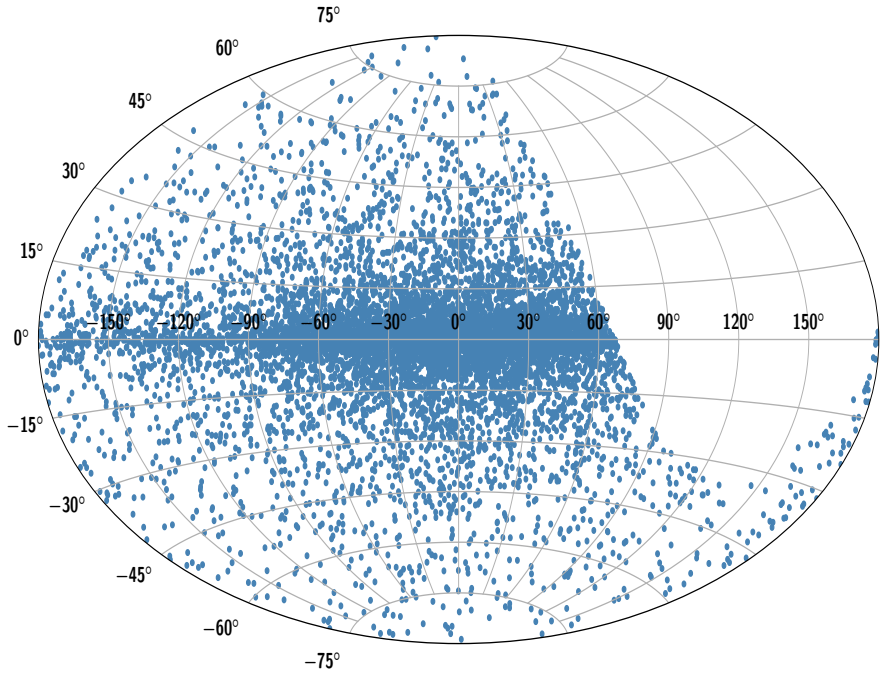}
	\caption{The Galactic distribution (Aitoff projection) of simulated normal pulsars that are potentially detectable by the considered hypothetical SKA-MID pulsar survey from a single realisation of Model-D.}
	\label{fig:SKA_GLGB}
\end{figure*}

We also note the dramatic increase in the number of pulsars discovered by the SKA-MID pulsar survey is accompanied by a much greater and wholesome sampling of its visible sky-region, as seen from Figure \ref{fig:SKA_XY} where we show the simulated pulsars projected in the Galactic plane, as well as in Figure \ref{fig:SKA_GLGB} where we show the Aitoff sky projection of the Galactic longitude-latitude of the simulated pulsars. Figure \ref{fig:SKA_GLGB} demonstrate the fact that  SKA-MID will be the main tool for discovering more distant pulsars in the Galactic plane ($|b|$ $\leq$ 10$^{\circ}$). 

\subsection{Prospects of detection of continuous gravitational waves }
\label{subsec:SKA_cgw}

The recent observations of gravitational waves from the inspiral and merger of binary compact objects heralded the era of gravitational-wave astronomy \citep{LIGO1, LIGO2}. Such extra-Galactic and transient events are, however, not the only potential sources of observable gravitational waves. One local source of gravitational waves that would provide a continuous signal is a rapidly rotating neutron star. Spinning neutron stars (isolated as well as those in binary systems) could generate detectable gravitational waves via several possible mechanisms. 

We consider the simplest and the most optimistic scenario for the case of an isolated neutron star, which we consider as a rigid star (with a triaxial moment of inertia ellipsoid) rotating around its principal axis of inertia and emitting gravitational waves from the $l = m = 2$ (spherical harmonics) mass quadrupole mode. In this case, the gravitational wave frequency $f_{gw}$ is expected to be twice the rotational spin frequency $f_{rot}$ ($=P^{-1}$) of the pulsar. 

Equating the gravitational-wave luminosity to the total kinetic energy lost as the pulsar spins down, we can obtain the so-called spin-down limit of the gravitational-wave strain ($h^{sd}_0$), which is given as \citep{CGW_Pulsars}:

\begin{equation}
    h^{sd}_0 = \Bigg( \dfrac{5 G I_{zz} \dot{f}_{rot}}{2c^3 d^2 f_{rot}} \Bigg) ^ {1/2} ~,
    \label{eq:GWspindown0}
\end{equation}
or equivalently, we can write
\begin{equation}
    h^{sd}_0 = 8.06 \times 10 ^{-19} \Bigg( \dfrac{|\dot{f}_{rot}|}{f_{rot}} \Bigg)^{1/2} \dfrac{I_{45}^{1/2}}{d_{\rm kpc}} ~,
    \label{eq:GWspindown}
\end{equation} where $I_{45}$ is the star's moment of inertia in the units of 10$^{45}$ ${\rm g~cm^{2}}$, and $d_{\rm kpc}$ is the distance to the pulsar in kiloparsecs. \\

In reality, especially for radio pulsars, only some fraction ($\eta$) of the spin-down energy gets converted to the gravitational radiation, and we have the gravitational wave strain to be $h_0$ = $\eta$ $h^{sd}_0$ \citep{CGW_Pulsars}. Since $\eta$ $<$ 1, it implies that $h_0$ $<$ $h^{sd}_0$. Therefore, it is reasonable to expect the detection of continuous gravitational waves only when the detection sensitivity reaches well below $h^{sd}_0$. Even if there is a non-detection of continuous gravitational waves when our detector's sensitivity has beaten the spin-down limit $h^{sd}_0$ of some particular pulsars, it will allow us to constrain the fraction of spin-down energy that can be attributed to the emission of gravitational-waves and we can also place stringent limits on the equatorial fiducial ellipticity of the star, which will be ultimately helpful for studies of the equation of state of the neutron star matter. So far, there are 20 young pulsars \citep{CGW_Pulsars} for which the sensitivity of the LIGO search runs have been able to beat their respective spin-down limits, but no detection of continuous gravitational waves have been claimed yet. We expect that future pulsar surveys and their precision timing campaigns will discover more pulsars with $f_{gw}$ in the range of ground based gravitational waves detectors as well as large enough values of $h^{sd}_0$, that are well inside the instrument sensitivity limits. 

Using a single realisation of Model-D, we extract the set of the synthetic pulsars that the proposed SKA-MID all-sky survey (Table \ref{tb:SKAsurveyList}) would be able to detect. For each such SKA-detectable pulsar, assuming the canonical value of the moment of inertia (i.e. $I_{45}$ = 1), we calculate their respective spin-down limits using equation (\ref{eq:GWspindown}) and check whether we will be in a position to detect continuous gravitational waves from them using more sensitive future LIGO observing science runs (LIGO A$+$).

\begin{figure*}
	\centering
	\includegraphics[width=0.80\textwidth,height=0.45\textheight]{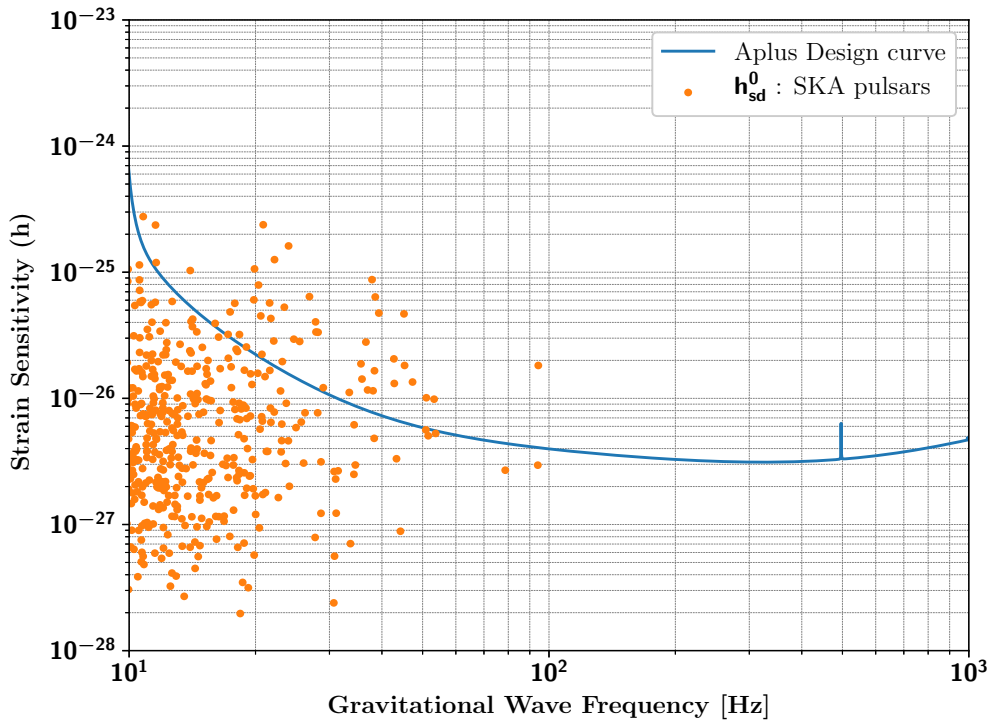}
	\caption{The spin-down limits on $h_{0}$ for simulated normal pulsars that are potentially detectable by the considered hypothetical SKA-MID pulsar survey, in a single realisation of model-D. The turquoise-blue solid line shows an estimate of the expected sensitivity of potential future science runs using LIGO A$+$.}
	\label{fig:SKA_CGW}
\end{figure*}

In Figure \ref{fig:SKA_CGW}, we show the spin-down limits $h^{sd}_{0}$ for the simulated normal pulsars that are potentially detectable by the considered hypothetical SKA-MID pulsar survey, in a single realisation of Model-D. The turquoise-blue solid line shows an estimate of the expected sensitivity of potential future science runs using LIGO A$+$. The search strain sensitivity is computed from the power spectral density ($S_n$) using the relation $h_{n}~=~10.4~\sqrt{S_{n}/T}$ assuming a typical observation time ($T$) of 1 year \citep{CGW_Pulsars}. For this purpose, we make use of the design data provided at the LIGO Document Control Center Portal\footnote{\url{https://dcc.ligo.org/LIGO-T1800042/public}}. From Figure \ref{fig:SKA_CGW}, we find that there will be a considerable number of SKA detectable normal pulsars emitting gravitational waves around the lower end of the LIGO A+ operational range of frequencies, and for about 50 of them, the value of $h_0^{sd}$ would be above the sensitivity limit ($h_n$).

\section{Summary and Conclusions}
\label{sec:conclusion}

In this work, we have revisited the population of normal pulsars in the Galactic field under the `evolutionary' approach. By comparing the distributions of various parameters of synthetic pulsars detectable by the Parkes Multibeam Pulsar Survey, the Pulsar Arecibo L-band Feed Array Survey, and two Swinburne Multibeam surveys with those of the real pulsars detected by the same surveys, we find that a good and physically realistic model can be obtained by using a uniform distribution of the braking index in the range of 2.5 to 3.0, a uniform distribution of the cosine of the angle between the spin and the magnetic axis in the range of 0 to 1, a log-normal birth distribution of the surface dipolar magnetic field (in Gauss) with the mean and the standard deviation as 12.85 and 0.55 respectively while keeping the distributions of other parameters unchanged from the population model \citep{fk06} most commonly used in the literature. We have replaced the universal `death-line' by a `death-condition' specific to each individual pulsar, depending on its braking index and inclination angle. With our best model, which we call as model-D, we predict that a prospective pulsar survey with phase-I SKA-MID will detect about nine thousand normal pulsars in the Galactic field. A good number of pulsars among these set will produce continuous gravitational waves in the operational range of the future ground-based gravitational waves detectors like LIGO A$+$, and about fifty of them will be promising candidates for the detection of continuous gravitational waves. We also provide a fit for the present-day distributions of the spin period and 1400 MHz luminosity of the whole normal pulsar population in the Galactic field, which are radio loud and beaming towards the earth, i.e. potentially observable with a sensitive enough telescope. From our analysis, we conclude that the underlying $L_{1400}$ distribution is well fit by a log-normal function (in the logarithm to the base 10) with mean $\mu_{log L} =-1.219$ ($L$ = $0.06$ mJy kpc$^2$) and standard deviation $\sigma_{log L}$ = $0.93$. The spin period distribution (in units of milliseconds) is also adequately described by a log-normal distribution (in the logarithm to the base 10) with $\mu_{logP} = 3.07$ and $\sigma_{logP} = 0.35$. These distribution functions can be readily used in future population studies under the snapshot approach.

\section*{Acknowledgements}

The authors thank Sam Bates for making PsrPopPy publicly available and replying to many queries regarding its usage. We also acknowledge the use of the cluster computing facility at High Performance Computing Environment (HPCE), Indian Institute of Technology Madras, Chennai, India, where the simulations were carried out. AC also thanks the HPCE staff members at IIT Madras for keeping the VIRGO super-cluster working amidst the Covid-19 pandemic and months-long national lockdown.

\section*{Data Availability}

This work has made extensive use of the population synthesis package PsrPopPy, which is publicaly available on Github \footnote{\url{https://github.com/samb8s/PsrPopPy}} and the data of real pulsars archived from the ATNF Pulsar Catalogue v1.63 \footnote{\url{https://www.atnf.csiro.au/research/pulsar/psrcat/catalogueHistory.html}}. The inputs used in running the simulations using PsrPopPy have been elaborately described in the text. The simulation results underlying this article, however, could be shared on reasonable request to the authors.

\bibliographystyle{mnras}
\bibliography{referenceList.bib} 

\appendix

\section{Some comments on PsrPopPy}

While working on this project, we discovered certain problems in the earlier version of the `evolve' module in PsrPopPy that have been reported to the developer for amendments. The following list of `alerts' are not exhaustive and concern only the functionalities of PsrPopPy that are used in the present work.

\begin{itemize}

\item The spiral structure of the Galaxy in the birth location of pulsars was not being generated correctly because in line 355 of the program {\tt galacticops.py}, $\sin \theta$ was erroneously written as $\cos \theta$.

\item In line 349 of {\tt galacticops.py}, while blurring the radial position, the standard deviation of the zero-centered Gaussian distribution  was set as `$0.50\times r$', which is different from that mentioned in \citet{fk06}, viz., `$0.07\times r$' for the YK2004 radial distribution model. We used `$0.07\times r$'.

\item There was an angle conversion error (between radians and degrees) while computing sine and cosine in the random alignment model, in lines 449 and 451 of the {\tt evolve.py} program. 

\item In the default settings of PsrPopPy, a distribution of $n$ is taken, yet the death line equation used was the one derived for $n=3$, i.e., equation (\ref{eq:BPdeathline}).

\item \citet{fk06} used an exponential distribution for the one-dimensional birth velocity components. However, in the default settings of the `evolve' interface in PsrPopPy, the one-dimensional birth velocity components were drawn using a Gaussian distribution centered at 0 km/s with width 180 km/s. We preferred to use the exponential velocity model proposed by \cite{fk06} in all our simulations because this model, owing to its heavier tail, can better accommodate for real pulsars with high one-dimensional velocities (see \cite{fk06} for examples). 

\end{itemize}

\bsp	
\label{lastpage}
\end{document}